\documentclass[12pt]{iopart}

\usepackage{graphicx}
\usepackage{epsfig}
\usepackage{amssymb}
\usepackage[numbers, sort&compress]{natbib}		

\begin{document}

\newcommand{\shorttitle}{Controlling fields in a microfabricated ion trap through design and compensation} 
\title[\shorttitle]{Controlling trapping potentials and stray electric fields in a microfabricated ion trap through design and compensation}

\author{S. Charles Doret\footnote{Author to whom any correspondence should be addressed.}, Jason M. Amini, Kenneth Wright, Curtis Volin, Tyler Killian\footnote{Present address: Ierus Technologies,  2904 Westcorp Blvd., Suite 210, Huntsville, AL 35805}, Arkadas Ozakin, Douglas Denison, Harley Hayden, C.-S. Pai, Richart E. Slusher, and Alexa W. Harter}
\address{
Georgia Tech Research Institute, Atlanta, Georgia, 30332\\
}
\ead{charlie.doret@gtri.gatech.edu}
\date{\today}

\begin{abstract}
Recent advances in quantum information processing with trapped ions have demonstrated the need for new ion trap architectures capable of holding and manipulating chains of many ($>10$) ions.  Here we present the design and detailed characterization of a new linear trap, microfabricated with scalable complementary metal-oxide-semiconductor (CMOS) techniques, that is well-suited to this challenge.  Forty-four individually controlled DC electrodes provide the many degrees of freedom required to construct anharmonic potential wells, shuttle ions, merge and split ion chains, precisely tune secular mode frequencies, and adjust the orientation of trap axes.  Microfabricated capacitors on DC electrodes suppress radio-frequency pickup and excess micromotion, while a top-level ground layer simplifies modeling of electric fields and protects trap structures underneath.  A localized aperture in the substrate provides access to the trapping region from an oven below, permitting deterministic loading of particular isotopic/elemental sequences via species-selective photoionization.  The shapes of the aperture and radio-frequency electrodes are optimized to minimize perturbation of the trapping pseudopotential.  Laboratory experiments verify simulated potentials and characterize trapping lifetimes, stray electric fields, and ion heating rates, while measurement and cancellation of spatially-varying stray electric fields permits the formation of nearly-equally spaced ion chains.  

\end{abstract}

\maketitle

\tableofcontents

\renewcommand{\leftmark}{\shorttitle} 

\section{Introduction}
Research with trapped atomic ions has progressed rapidly in recent years~\cite{WL11, DM10, HRB08}, driven in large part by interest in developing a large-scale quantum information processor.  Much recent work has focused on scaling experiments to larger numbers of ions, with particular developments in multi-ion entanglement~\cite{HHR05e, LKS05e} and quantum simulation~\cite{KCK10e, EKK10e, IEK11e}.  One plan for working with large numbers of ions utilizes anharmonic trapping potentials to stabilize chains against structural instabilities~\cite{LZI09e}.  An alternative approach involves interconnected networks of many trapping zones~\cite{KMW02, WMI98e}.  Both approaches require the use of many trap electrodes; as experiments move toward larger scales, it is imperative that ion trap technologies keep pace.  

One technology that is particularly amenable to scaling and complex geometries is the surface-electrode trap~\cite{CBB05e, SCR06e}, where all of the electrodes are in a common plane.  This geometry is easily miniaturized, which is advantageous for scaling both because more electrodes may be fit into a given area and because multi-qubit gate-operation times scale favorably with decreasing trap dimensions.  Planar geometries are also amenable to microfabrication as a monolithic structure, eliminating the challenges of aligning and assembling the components of (more traditional) multi-substrate or macroscopic four-rod Paul traps.  Microfabrication also enables simultaneous construction of many virtually identical traps, and should permit assembly of combinations of linear sections with junction elements~\cite{AUW10e, MHS11e, WAF12e} to form interconnected networks of many trapping zones.  

In this work we present a microfabricated trap that incorporates three features which improve the fine control of trapping potentials: microfabricated on-chip capacitors that locally suppress radio-frequency (RF) pickup on DC electrodes, a top-level ground layer covering 97\% of the chip surface that simplifies modeling of trap potentials by shielding electrode leads, and a loading aperture which, along with locally shaped RF rails, reduces deformation to the RF pseudopotential tube in which ions are confined.  Traps are fabricated on silicon substrates from patterned layers of aluminum and silicon dioxide (SiO$_{2}$) using standard CMOS-compatible techniques similar to those described in~\cite{LLC09e}.  We have used several of these traps to study lifetimes and heating rates of single ions and to create  single- and dual-species ion chains.  Using a single ion as a probe, we experimentally characterize the spatial and temporal dependence of stray electric fields along the length of the trap.  Treating the measured fields as model inputs, we derive global nulling potentials that cancel stray fields throughout the trap, enabling the formation of nearly-equally spaced ion chains in anharmonic potential wells.   

\section{Trap design}
\label{sec:design}

\begin{figure}
\center
\includegraphics[width=12.0 cm]{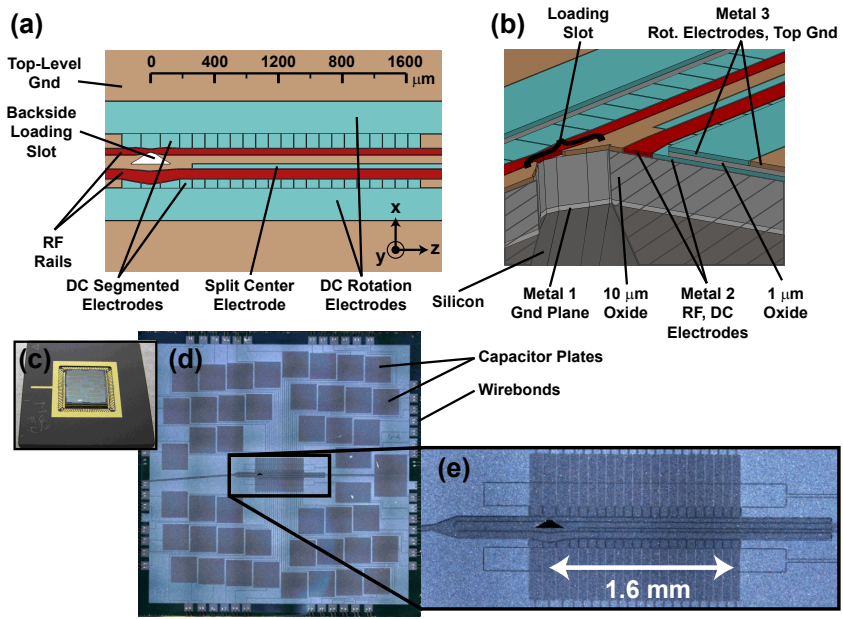}
\caption{\label{fig:trap} (a) A schematic view of the active region, and (b) a cutaway view showing internal layers (vertical direction is 10$\times$ expanded to make thin internal layers visible).  (c) A packaged trap mounted in a 100 pin CPGA.  (d) Dark-field optical microscope image of the trap chip.  Note the arrangement of the wire bonds, which permit optical access at 90$^{\circ}$ and 45$^{\circ}$. (e) An enlargement of the active region.  The darker shade indicates components fabricated in the middle metal layer, including RF rails, DC electrodes, and on-chip capacitors.  Components are effectively visible even where covered by the top metal layer due to layer print-through and slight differences in surface roughness.}
\end{figure}

The trap is based on an asymmetric five-wire geometry~\cite{CBB05e} with a split center DC electrode and segmented outer DC electrodes (figure~\ref{fig:trap}).  Three metallic conducting layers are created from 99\%~Al~/~1\%~Si~\cite{AlCu} and are separated from one another by SiO$_2$.  The bottom metal layer forms a ground plane that prevents RF electric fields from penetrating into the RF-lossy p$^{++}$-doped silicon substrate.  A thick insulating layer of 10~$\mu$m of SiO$_2$ between the middle (electrode) layer and this ground plane limits the capacitance and power dissipation of the RF electrodes.  A second ground plane covering most of the trap is separated from the electrode layer by a 1~$\mu$m SiO$_2$ layer.  This top ground defines the outer boundary of the segmented control electrodes and forms the second plate for capacitors patterned into the electrode leads (see below).

The trap is designed around a target ion height of 63~$\mu$m~\cite{ionheight}.  DC and RF rail widths and locations are chosen using the conformal mapping described in~\cite{Wes08} to obtain this height while rotating the natural (DC) secular axes 25 degrees from the trap normal.  This axis rotation ensures that both radial secular modes may be Doppler-cooled by a single laser oriented along the trap surface.  RF rails 30 and 50~$\mu$m in width provide strong radial confinement at modest voltages (radial secular frequencies of $\sim$4~MHz at $\sim$100~V for $^{40}$Ca$^+$) while keeping power dissipation low.  The DC rails are segmented into control electrodes with a 79~$\mu$m pitch (except for those around the loading slot, which are 120~$\mu$m).  Additional DC electrodes that run along the entire length of the trap permit axis rotation and micromotion compensation over long chains of ions.  All gaps between electrodes are 4~$\mu$m wide near the ion, expanding to 10~$\mu$m under the top-level ground to reduce the probability of unintended shorts or RF breakdown.  Electrical connections are made via wire bonds strategically placed around the trap perimeter to provide optical access along key directions.

\subsection{Shaped loading slot}

A common failure mode for surface-electrode traps is shorting between electrodes due to neutral atom flux contaminating the trap surface during trap loading; such contamination can also negatively impact trap heating rates~\cite{DK02, TKK00e}.  One solution is to pierce the trap substrate with a slot, allowing the neutral atom source to be relocated behind the trap, so that the trap substrate prevents accumulation on the electrodes (first utilized in~\cite{LLC09e}).  These ``back-side loading slots" must necessarily lie directly below the pseudopotential null, causing variations in trap strength over the slot as well as pseudopotential barriers at the slot edges.  Loading slots also locally affect the ion height, requiring adjustments to laser heights as ions are shuttled into or away from the loading zone.  One resolution is to extend the loading slot along the entire trap~\cite{SFH10e}.  However, this increases sensitivity to stray fields from sources behind the chip, reduces ion lifetime while loading due to collisions between neutrals and trapped ions, and complicates deterministic loading of a particular number or sequence of ions.  Alternatively, spatially separating the loading slot from computational regions~\cite{AUW10e, MHS11e} restricts pseudopotential deformations from the slot to remote areas.  However, this strategy cannot be used if optical or other components~\cite{BEM10e} need to be integrated adjacent to computational regions.

\begin{figure}
\center
\includegraphics[width=12.0 cm]{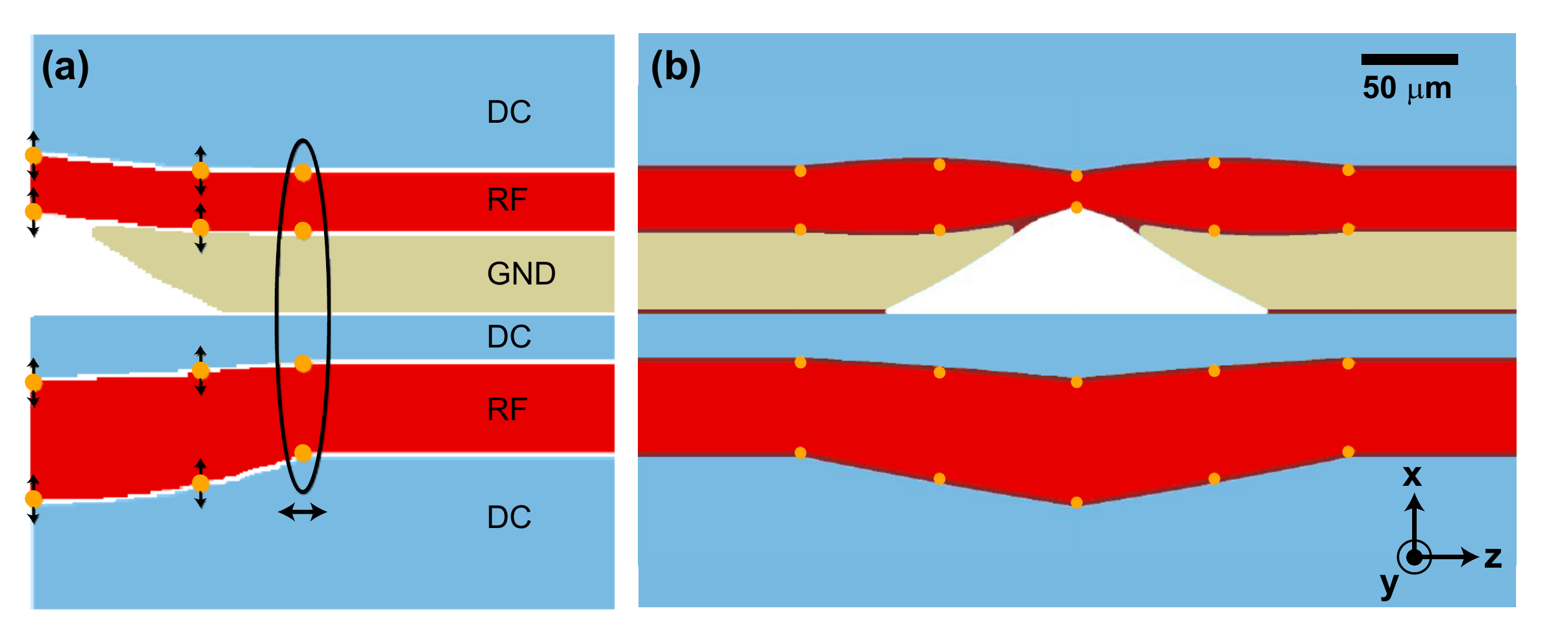}
\caption{\label{fig:loading_slot} Loading slot and RF rail optimization.  (a) Initial geometry and control points used as a starting point for the optimization.  Due to symmetry considerations only half of the slot need be simulated.  (b) Final optimized shape.}
\end{figure}

To address these issues it is desirable to minimize local pseudopotential deformations and maintain a constant ion height over the loading zone.  This may be accomplished by replacing the parallel RF rails with shaped rails to correct for the perturbation caused by a loading slot.  As a starting point, we define an initial geometry that qualitatively addresses the problem (figure~\ref{fig:loading_slot}).
First, a modified slot with tapered edges is introduced (here a trapezoid).  Second, the separation between the RF rails is increased around the slot, raising the height of the pseudopotential tube~\cite{Wes08}.  Third, the rail widths are increased, strengthening the radial confinement.  To optimize this modified geometry the shape is characterized in terms of a set of twelve variable ``control points" (nine degrees of freedom) that completely define the edges (figure~\ref{fig:loading_slot}a).  The locations of these points are systematically varied with a downhill-gradient search.  Overlap between each new geometry's electric field and a target field serves as a fitness function to measure progress towards a solution.  Fields are calculated with an in-house boundary element method (BEM) electrostatics solver capable of solving problems with more than $10^{6}$ unknowns, similar to those described in~\cite{SPM10e, BOV11e}.

\begin{figure}
\center
\includegraphics[width=12.0 cm]{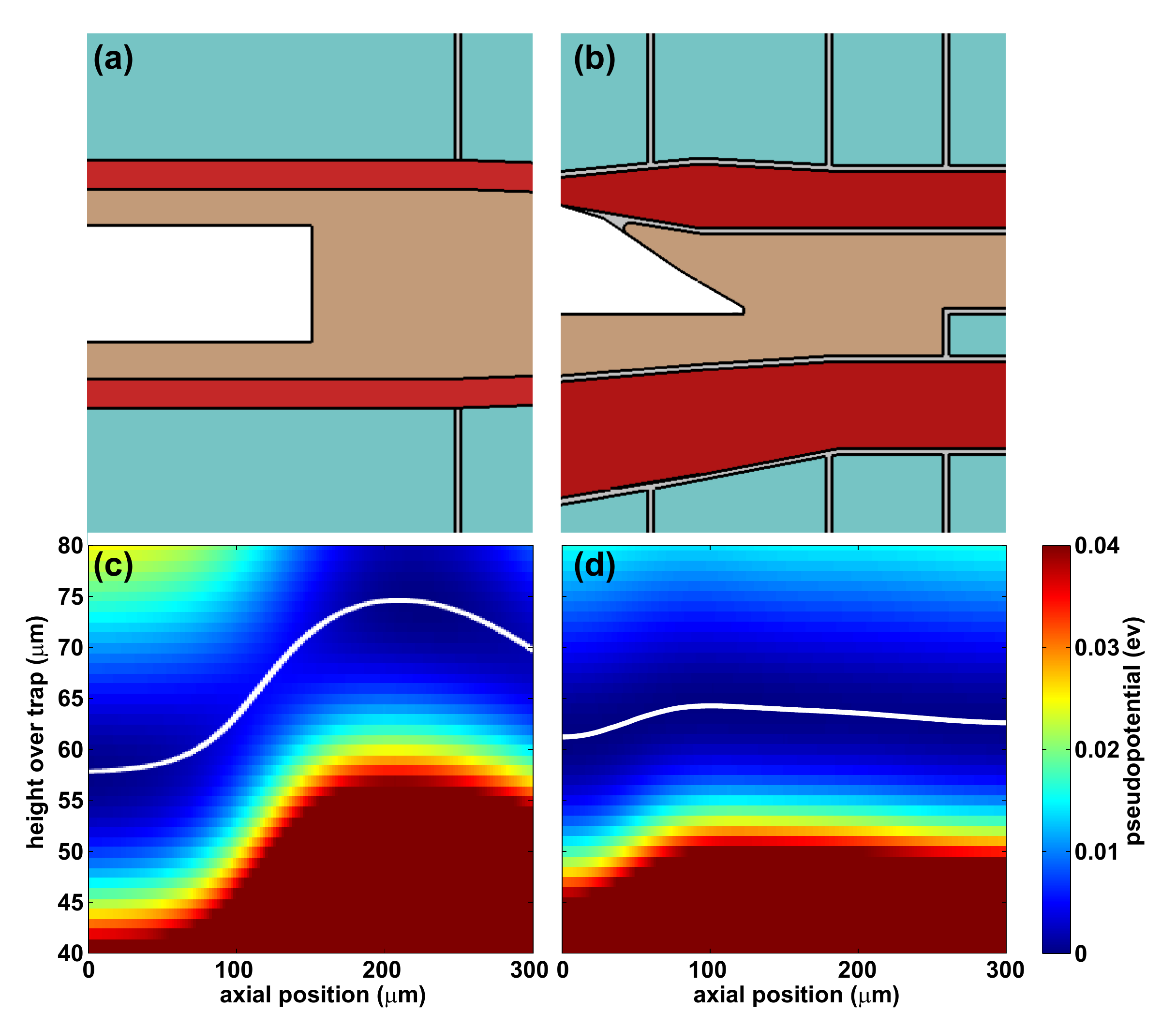}
\caption{\label{fig:loading_slot2} Back-side loading slot comparison.  (a) The slot geometry in~\cite{LLC09e}, and (b) the modified slot geometry.  (c,d) Pseudopotential resulting from (a) and (b), respectively; horizontal scale is shared throughout.  Simulations are for 60~MHz RF drive frequency and 3.5~MHz radial secular frequencies.  The white line represents the height of the RF-null.  Note especially the reduced variation in ion height at the edge of the load zone in the optimized geometry.}
\end{figure}

Using this procedure we find an improved slot geometry that reduces variations in ion height from the 17~$\mu$m resulting from the rectangular slot in the trap described in~\cite{LLC09e} to less than 3~$\mu$m (figure~\ref{fig:loading_slot2}).
The optimized geometry largely adheres to the qualitative model described above, with the exception of the central ``pinch" in the narrower RF rail which compensates for a bend in the pseudopotential tube in the $\hat{x}$-direction that would otherwise result from the asymmetry of the loading slot.

\subsection{Top-level ground}

A major difficulty for the design and use of segmented ion traps is the challenge of accurately simulating the electric fields produced by complicated electrode geometries.  BEM solvers, commonly used to simulate trapping potentials, work by discretizing (meshing) the electrode surfaces into small elements (often triangles) and then solving for the charge density on the electrodes and the associated fields.  Since the charge density exhibits a singularity along electrode edges, a fine mesh is required in these areas; a coarser mesh is adequate away from edges.  Even with this ``adaptive meshing," complicated electrode geometries generally have tens or hundreds of thousands of elements. 

To reduce the problem size (and the corresponding requirements for computation time and computer memory), a conventional approach is to restrict the modeling to the immediate vicinity of the ion.  While this is a reasonable approximation in some situations, increased modeling accuracy requires the inclusion of more distant pieces of the trap geometry.  Such accuracy is especially important when calculating transport waveforms (sets of smoothly varying potentials) that merge and separate ions, because these waveforms balance the fields from many electrodes, magnifying small errors.  Unfortunately, distant areas generally contain electrode leads and other small features, requiring fine meshing and dramatically increasing problem size.  A better solution to improve modeling accuracy is to shield these distant areas with an additional ground plane above the electrode layer.  This replaces the many edges of individual electrode leads with a single large feature, most of which may be coarsely meshed.  An added ground plane also improves the match between modeling and experiment by shielding the ion from insulating trap features that might otherwise be at an undetermined potential.  Additionally, the top metal layer protects underlying structures from scratches incurred during wafer dicing and trap packaging, increasing yield.    

\subsection{On-chip capacitors}

The top-level ground may be used in combination with the middle metal layer to form capacitive filters for the DC electrodes.  Such capacitors are an essential element for reducing RF pickup on these control electrodes.  Placing the capacitors close to the trap is important, because their effectiveness is impaired by any intervening resistance or inductance, such as that of a lead between the trap and a vacuum feedthrough~\cite{AHJ11e}.  The trap described here incorporates 1~mm$^2$ capacitor plates into the leads for the DC control electrodes (see figure~\ref{fig:trap}d), providing 60~pF to ground.  Along with the stray capacitance between individual DC electrodes and the RF rails (calculated to be $\sim 1.4 \times 10^{-3}$~pF), these capacitors create a capacitive divider that shorts RF signals to ground, suppressing the RF pickup by a factor of approximately $4.3~\times~10^{4}$.

\section{Fabrication and packaging}
\label{sec:fabrication}

Trap fabrication uses standard silicon very-large-scale-integration (VLSI) processing steps~\cite{Jae01}.  To begin, we clean a p$^{++}$-doped silicon wafer (0.5~mm thickness) using a piranha etch followed by a buffered oxide etch, removing contaminants and surface oxidation to ensure good metal adhesion.  We then etch most of the way through the wafer from the back side with potassium hydroxide (KOH) to form a tapered rectangular relief that will later become the loading slot.  A 1~$\mu$m layer of aluminum, bonded to the wafer with a 30~nm titanium underlayer, is deposited on the front side to form the bottom ground plane.  This layer is lithographically patterned and plasma-etched to define the trap loading slot and alignment marks for ensuing layers.  We then deposit a thick (10~$\mu$m) film of low stress oxide ($\sim$20 MPa for a 10~$\mu$m layer) via plasma-enhanced chemical vapor deposition (PECVD) to separate the electrode layer from the bottom ground plane; 10~$\mu$m is adequate to reduce the RF-electrode capacitance to $\sim$1~pF.  A second aluminum layer (also bonded with a 30~nm titanium adhesion layer) is deposited on top of this oxide.  This layer contains essential trapping structures, including RF and DC electrodes, electrode leads, capacitor plates, and wire bond pads; it is also patterned with photolithography and plasma-etching. 

We then deposit the top-level ground, a 1~$\mu$m aluminum layer (again bonded by a 30~nm titanium underlayer) over 1~$\mu$m of  oxide.  A deep inductively-coupled plasma (ICP) oxide etch removes oxide exposed in the gaps between electrodes.  To complete the loading slot an ICP-Bosch etch from the back side of the wafer removes the remaining substrate material in the KOH etched region to open the loading slot aperture.  We coat the back side with a titanium adhesion layer and 300~nm of gold to prepare the chips for ultra-high vacuum (UHV) compatible mounting.  The wafer is then cleaned and diced into chips.  Final die-level oxygen plasma and argon plasma treatments prepare the trap surface for wire bonding.  

To package the traps for vacuum installation, chips are mounted to a ceramic pin grid array (CPGA) carrier via a 1.2~mm tall slotted alumina spacer with 80/20 Au/Sn solder (figure~\ref{fig:trap}c).  The spacer raises the top of the chip above the CPGA perimeter to permit access for laser beams parallel to the trap surface.  Aluminum wires (25~$\mu$m diameter) are wedge bonded to the chip and carrier to make all electrical connections with two wires used on each connection for redundancy.  The resulting carrier/trap package is ready for in-vacuum installation.

\section{Experimental characterization and control}

To date we have trapped ions in ten of these traps.  We have extensively characterized five, measuring stray electric fields and secular mode frequencies, determining heating rates, and quantifying ion loss during ion transport.  We have built single-species chains of up to twelve ions as well as shorter multi-species chains.  All data presented here were collected with $^{40}$Ca$^+$ and/or $^{44}$Ca$^+$.  As calcium trapping has been reviewed elsewhere~\cite{HRB08}, we confine discussion of our particular experimental setup to the appendix. 

\subsection{Ion shuttling and model verification}
\label{sec:shuttling}
Ions may be moved between different points in the trap by applying a transport waveform to the DC control electrodes.  Waveforms are constructed by deriving potential sets that create an axial harmonic well at regularly spaced intervals (usually every 10~$\mu$m) throughout the trap.  Our waveforms typically are designed for an axial secular frequency $\omega_{z}=2\pi\times1$~MHz for $^{40}$Ca and a 12$^{\circ}$ rotation of the radial axes from the trap normal.  Intermediate waveform positions are generated by interpolating between adjacent wells to produce smooth transport.  

Sequentially updating DC control voltages according to the appropriate transport waveform generates a moving potential well that can shuttle an ion between different locations.  Waveforms for merging and splitting of multiple ions between adjacent potential wells may be generated in a similar fashion from suitable superpositions of wells when ion-ion repulsion is included in the model.  Typical waveforms utilize 1000 steps for transport across the entire 1.4~mm length of the trap, giving a maximum ion velocity of 0.7~m/s at 500~kHz DAC update rate.  Modeled potential wells show excellent agreement with their implementation in fabricated traps.  We characterize axial harmonic wells by measuring secular frequencies and ion positions at 10~$\mu$m intervals, finding agreement with simulations at the 3\% level throughout the entire trap excepting the loading zone.  We find that loading rates and agreement with simulation in the vicinity of the load zone are sometimes improved by locally perturbing the potentials to counteract stray electric fields near the slot.  These fields may arise from sources behind the trap and penetrate the slot, or from stray charge nearby.  Stray charge near the loading slot is thought to originate from poor collimation of the UV photoionization laser~\cite{WLL11e}.

\subsection{Stray field measurement and global compensation}
\label{sec:fields}

The close proximity of the lasers to the trap in small surface-electrode traps can lead to the accumulation of stray charges~\cite{WLL11e, HBH10e}.  These charges create fields that push ions away from the pseudopotential null, sometimes distorting the potential enough to prevent trapping.  Uncompensated stray fields cause excess micromotion which negatively impacts Doppler cooling, fluorescence collection, and ion heating rates~\cite{LBM03e}.  They also affect the spacing and stability of ion chains and may contribute to ion loss during shuttling operations.  Detailed measurement and characterization of stray fields is therefore essential for optimizing trap operation.

\begin{figure}
\center
\includegraphics[width=12.0 cm]{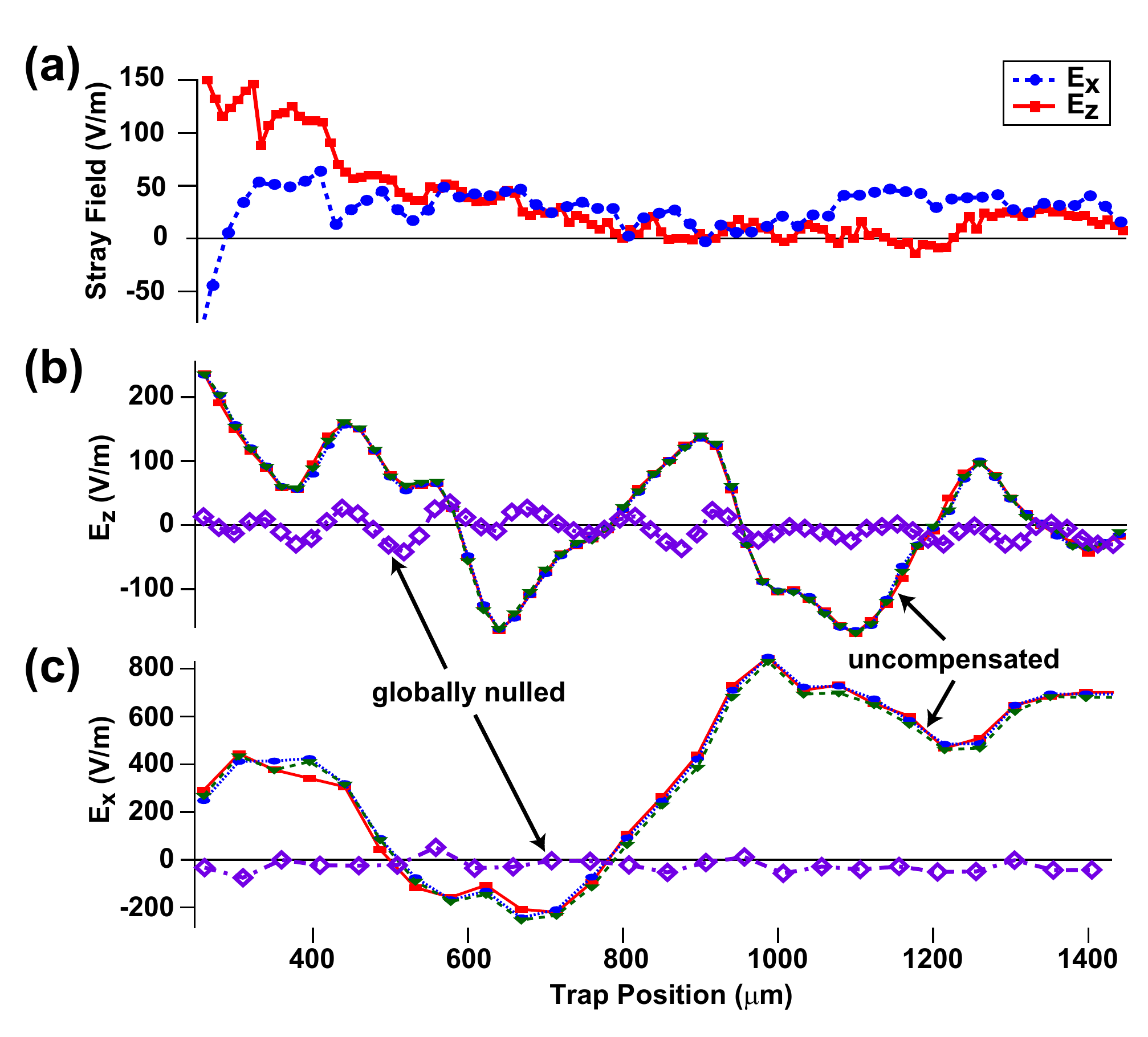}
\caption{\label{fig:stray_fields} Stray field stability and compensation.  
(a) $E_{x}$ and $E_{z}$ as measured in a typical trap.  (b) $E_{z}$ and (c) $E_{x}$ in a trap with unusually large stray fields, measured on three consecutive days after initial transient charging behavior had saturated.  Stray fields are simultaneously cancelled to less than 50~V/m throughout the entire trap ($\diamondsuit$) by applying a global nulling potential.  Typical error-bars are 10 and 20 V/m for $E_{z}$ and $E_{x}$, respectively.}
\end{figure}

Along the $\hat{z}$ direction the dominant effect of stray fields is to shift the location of the harmonic well created by the DC potentials.  For a stray field $E_{z}$, ion mass $m$, and axial secular frequency $\omega_{z}$, the shift is given by $z_{ion} = eE_{z}/mw_{z}^{2}$.  Scaling the strength $\omega_{z}$ of the axial well while simultaneously monitoring the position of an ion allows us to extract the magnitude of the stray field.  By automating this procedure and measuring at many ion locations, we produce a detailed map of $E_{z}$ (figure~\ref{fig:stray_fields}a).  Measurement precision is approximately 10~V/m, limited by the effective size of our camera pixels (1.6~$\mu$m).

To measure stray fields along $\hat{x}$~\cite{axes} we follow a procedure similar to that outlined in~\cite{PAB99e}: the (resolved) micromotion sidebands on the 397~nm cooling transition are minimized by applying a known field via the DC electrodes.  This determines the stray field with a precision of approximately 10-20~V/m (figure~\ref{fig:stray_fields}a).  In principle, stray fields along $\hat{y}$ could be measured in the same way.  Unfortunately, lasers aligned in the $\hat{x}-\hat{z}$ plane are insensitive to micromotion along $\hat{y}$, so alternative methods~\cite{ASS10e, NDM11e} are more practical.  As fields along $\hat{y}$ do not strongly impact Doppler cooling or shuttling fidelity in this trap, we do not map them exhaustively in general.  However, we can indirectly extract $E_{y}$ from the splitting between the radial secular modes, which depend on $E_{y}$ due to anharmonicities in the RF-pseudopotential.  Comparison of measured and modeled splittings allows us to estimate $E_{y}$, which is usually less than 200~V/m.   

We find that stray fields vary at the level of approximately 100~V/m/day during the first few days of trap operation, possibly due to laser-induced charging of surface contaminants and oxide.  This variation saturates after approximately one week, after which stray fields vary at the level of a few tens of V/m or less (figure~\ref{fig:stray_fields}b,c).  After saturation the secular mode frequencies are stable to one part in $10^{-4}$ over hours and one part in $10^{-3}$ over days during normal trap operation. 

We use the measured stray field profile as a model input to calculate a trap-specific field-nulling potential that simultaneously cancels these fields along nearly the entire trap.  Nulling solutions are calculated from the BEM model by varying the applied potentials on all segmented electrodes in a downhill-gradient search to match the measured $E_{z}$ and $E_{x}$ while maintaining $E_{y}=0$.  In addition to matching the measured fields, the optimization minimizes $\delta E_{z}/\delta x$ and $\delta E_{z}/\delta y$ to prevent the secular axes from being skewed relative to the pseudopotential symmetry axis.  Two additional first order parameters remain unconstrained, corresponding to a radial quadrupole field rotated at an arbitrary angle.  However, this causes only slight variations in radial secular axis directions and frequencies for which we may account with further experimental characterization.  Nulling solutions generally require only modest ($\sim \pm1$~V) voltages and cancel stray fields to less than 50~V/m (figure~\ref{fig:stray_fields}b,c).

\subsection{Trapping lifetime and ion chains}
\label{sec:chains}

\begin{figure}
\center
\includegraphics[width=12.0 cm]{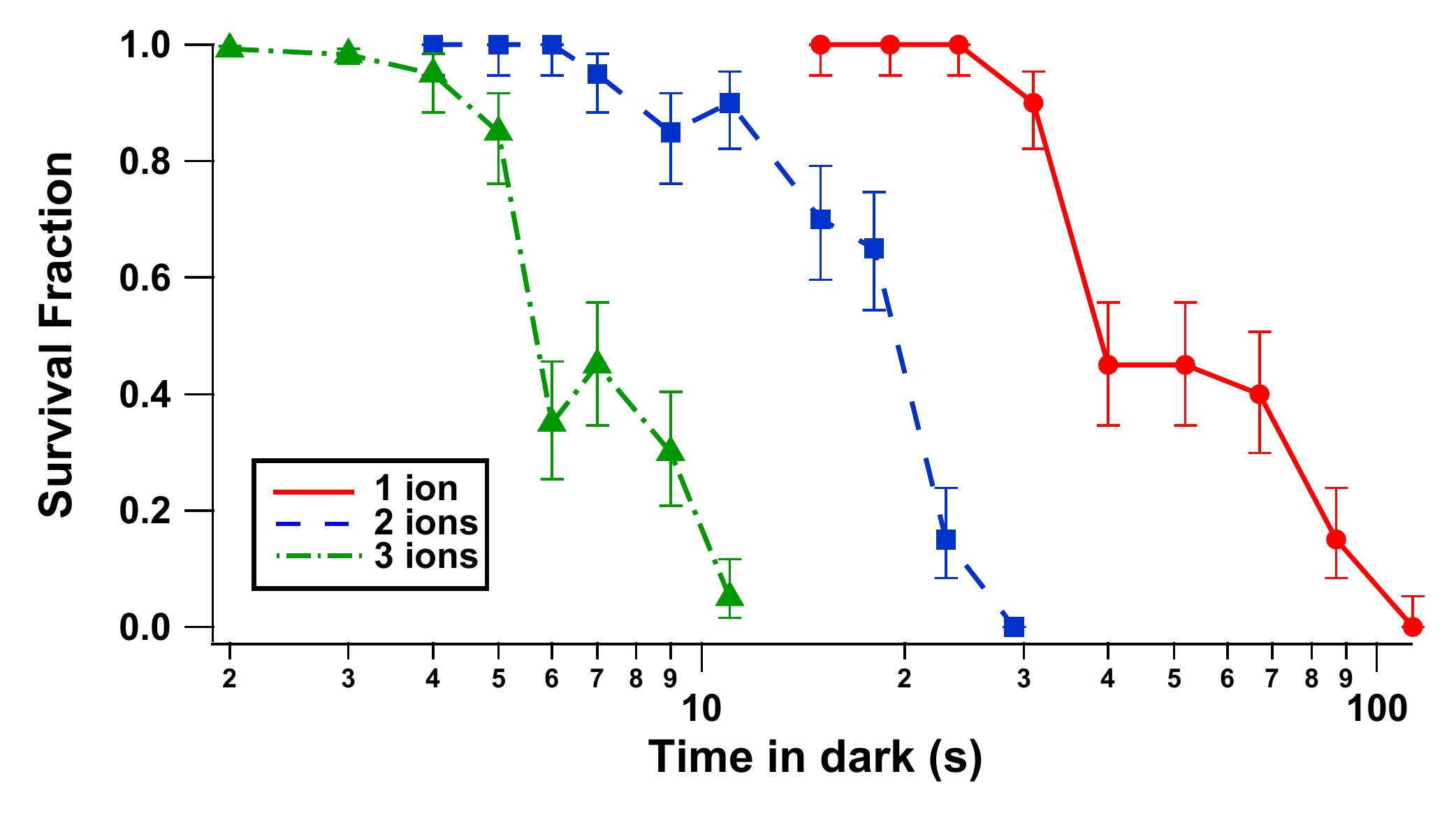}
\caption{\label{fig:dark_lifetimes} Lifetimes without Doppler cooling for single ions and small chains trapped in a harmonic ($\omega_{z} = 1$~MHz) well.  Survival is defined as the entire chain remaining intact; no distinction is made between the loss of one or several ions.}
\end{figure}

The trapping lifetime of a single Doppler-cooled ion is many hours or longer.  The lifetime without cooling is approximately 45 seconds and is strongly non-exponential: losses are negligible for the first 25 seconds after cooling is stopped.  Chains of two or three ions display similar loss characteristics with shorter lifetimes (figure~\ref{fig:dark_lifetimes}).

\begin{figure}
\center
\includegraphics[width=12.0 cm]{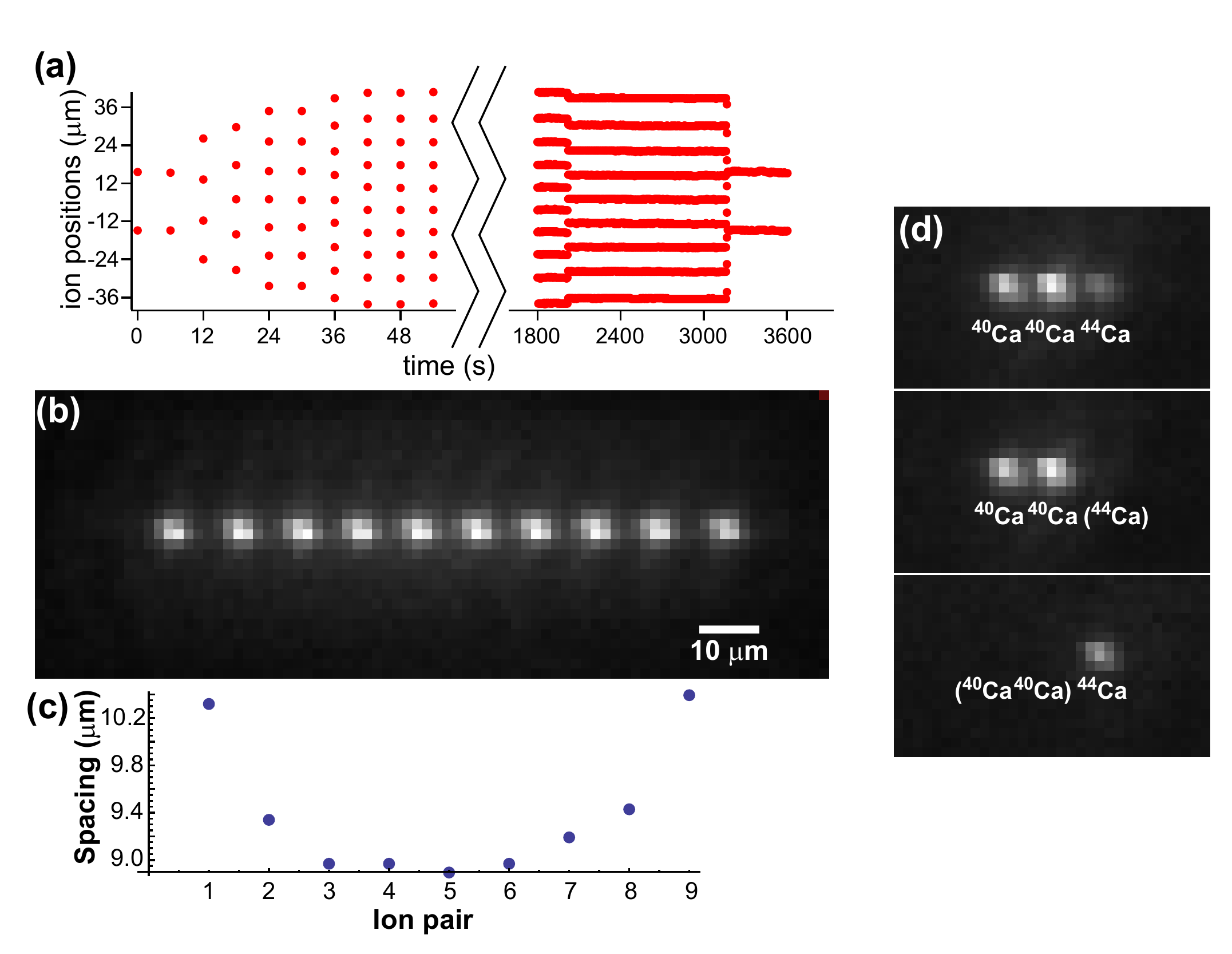}
\caption{\label{fig:chains} Ion chains.  (a) Ion-by-ion chain formation.  During the first 42~s individual ions are loaded into a well at the loading zone and shuttled to a chain stored 900~$\mu$m away.  After 42~s loading and shuttling is stopped and the chain is held in a static anharmonic potential well with continuous Doppler cooling.  This 10-ion chain was stored for approximately 30 minutes before ion loss.  (b) CCD image of a 10-ion chain.  (c) Spacings between ion pairs in (b), demonstrating less than $5\%$ variation in spacing for the central 8 ions (a factor of three times more uniform than for a harmonic well with the same central spacing).  Error-bars are comparable to the size of the points.  (d) A mixed-isotope chain.  (top) The chain is illuminated with two frequencies of Doppler cooling light.  (lower panels) Only one of the two frequencies is used.}
\end{figure}

An advantage of this trap geometry is the ability to deterministically load rare isotopes or mixed-species chains with a predetermined number and sequence of ions.  Whereas ions trapped in geometries with long, open loading slots are susceptible to charge exchange collisions during loading~\cite{LRH04e}, the small slot in our geometry localizes the neutral atom flux and significantly reduces the likelihood of collisions for ions trapped far from the slot.  By loading an ion in a potential well located above the loading slot, transporting it across the trap, and merging it with a second (static) well, it is possible to construct a chain of ions one ion at a time.  Laser frequencies for photoionization and Doppler cooling at the loading zone may be chosen independently for each ion loaded, making it possible to build a chain with any desired sequence of ions.  We utilize automated chain-loading routines that control laser frequencies and trapping potentials while analyzing camera images for ion positions in real-time so as to load chains of a particular size and isotopic sequence (figure~\ref{fig:chains}).
Mixed-isotope chains may be cooled with only a single wavelength~\cite{HMS09e} (i.e. sympathetic cooling).  We cool chains of $^{40}$Ca$^{+}$ and $^{44}$Ca$^{+}$ with a single laser tuned near the $^{40}$Ca$^{+}$ cooling transition.  However, we find that the 842~MHz isotope shift is too small to permit robust sympathetic cooling using $^{44}$Ca$^{+}$ as the ``coolant" ion since in this case the laser is blue detuned relative to $^{40}$Ca$^{+}$, causing Doppler-heating on that isotope.

Stability and lifetime of ion chains can be improved by utilizing a ``flat-bottomed" potential instead of a harmonic well.  Such anharmoinc potential wells are designed to hold ion chains with nearly-equal spacing between ions, mimicking the confinement that would be provided by placing a half-infinite chain of equally-spaced ions at each end of the trapped chain.  This potential shape allows long chains of many ions to be stored without transitioning to a zig-zag structure \cite{LZI09e} as more ions are added.  Forming these potentials requires the use of many ($>$~10) DC electrodes to strengthen the curvature at the edge of the well and compensate for the repulsive effect of the chain's space charge, but they provide greater trap depth and improved chain lifetime compared to harmonic wells~\cite{LZI09e}.  Spacing between ions in such wells is extremely sensitive to axial stray fields.  Any curvature in the axial field profile distorts the chain by clustering ions together or spreading them apart, and this effect is amplified as the number of ions grows.  Nevertheless, after cancelling stray fields (section~\ref{sec:fields}), we are able to construct ion chains with less than 5\% variation in spacing between all but the end ions (figure~\ref{fig:chains}b,c).

\subsection{Ion heating}
\label{sec:heating}

\begin{figure}
\center
\includegraphics[width=12.0 cm]{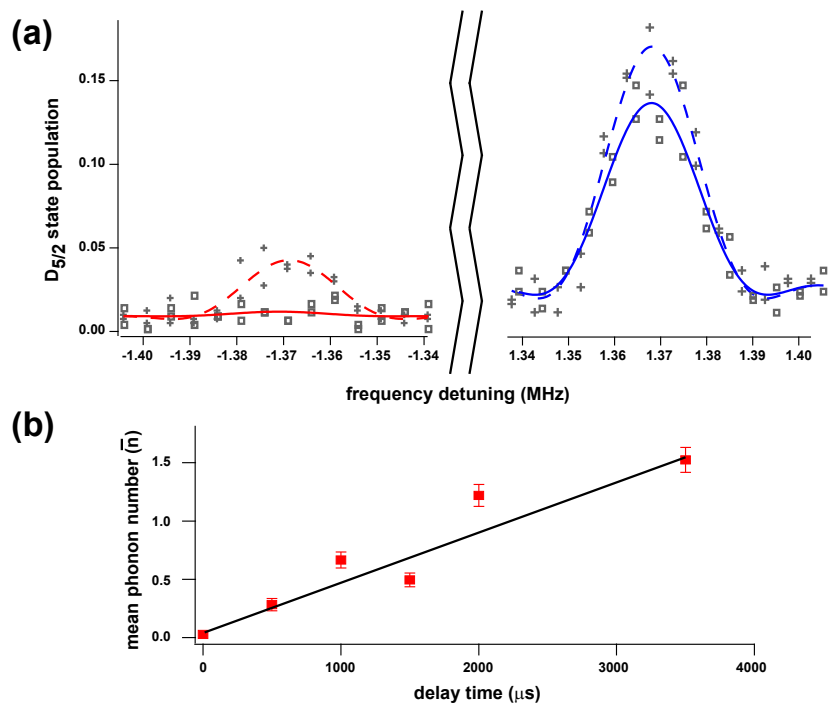}
\caption{\label{fig:heating} Ion heating.  (a) Axial motional sidebands on the 729~nm $S_{1/2} \rightarrow D_{5/2}$ transition.  Data indicated with squares ($\square$) is taken immediately following sideband cooling, while crosses ($+$) are after a delay of 0.5~ms.   (b) Heating of the axial secular mode ($\omega_{z} = 2\pi\times1.37$~MHz).  A linear fit yields a heating rate of 0.41(4) quanta/ms.}
\end{figure}

A critical parameter for any ion trap is the rate at which an ion is heated.  The largest source of heating in ion traps remains incompletely understood, but it seems to be related to contaminants on the trap surface and is generally much larger than that expected from Johnson noise~\cite{HCW11e}.  Following Turchette \emph{et al}.~\cite{TKK00e}, we determine the heating rate by monitoring the ratio of the motional sideband intensities on the calcium $^{2}$S$_{1/2}~\rightarrow~^{2}$D$_{5/2}$ transition at 729~nm (figure~\ref{fig:heating}).  We find a heating rate of 0.41(4) quanta/ms at an axial secular frequency of 1.37~MHz, corresponding to a spectral noise density $S(E) = 3.9(4) \times 10^{-12}~$V$^{2}/$m$^{2}$ Hz.  This is the lowest spectral noise density measured at room temperature to date for traps fabricated with CMOS techniques~\cite{HCW11e}, and it is comparable to reported heating rates in other traps of similar size.     

\section{Conclusions}
\label{sec:conclusions}

We have developed and characterized a surface-electrode ion trap fabricated with CMOS compatible techniques.    In contrast to previous CMOS traps, this trap adds a third, grounded metal layer that shields the ions from potentials on distant trap structures, thereby reducing computational requirements for accurate electric-field modeling.  By combining accurate simulations with precise, detailed measurements of stray electric fields, we generate DC potentials that null these stray fields simultaneously throughout large regions of the trap.  This makes possible the reliable merging of potential wells for deterministic loading of ion chains and the formation of equally-spaced ion chains in anharmonic potential wells.  Ion heating is low enough to incorporate this trap design into a future large-scale quantum information processor.  




\section*{Acknowledgments}
\addcontentsline{toc}{section}{Acknowledgments} 

We thank Kenton R. Brown for many helpful comments on the manuscript, and Chris Shappert for help with capacitance modeling.  SCD thanks the Georgia Tech Research Institute for postdoctoral fellowship funding.  This material is based upon work supported by the Office of the Director of National Intelligence (ODNI), Intelligence Advanced Research Projects Activity (IARPA,) under US Army Research Office (ARO) contracts W911NF081-0315 and W911NF101-0231. All statements of fact, opinion, or conclusions contained herein are those of the authors and should not be construed as representing the official views or policies of IARPA, the ODNI, or the U.S. Government.

\section*{Appendix: Experimental Details}
\addcontentsline{toc}{section}{Appendix: Experimental Details} 
\label{sec:experiment}

\subsection*{Vacuum installation}

The CPGA/trap package is installed into a 100-pin socket formed by sandwiching individual pin receptacles between two UHV compatible PEEK plates.  Each receptacle is crimped to a Kapton insulated wire, and the wires are bundled into four groups of 25 wires which terminate at commercial PEEK DB-25 connectors.  The socket is screwed to a stainless steel (SS) base and rigidly mounted inside a 4.5-inch spherical octagon (Kimball Physics).  A resistive oven beneath the socket produces neutral atoms directed through the loading slot.  A SS screen (100 lines-per-inch, 80\% optical transmission), attached above the carrier, shields the trap from stray electric fields.  Low resistance electrical feedthroughs on two side ports of the octagon are used to feed the calcium oven and the trap RF.  The top of the octagon is closed with a viewport for ion imaging and fluorescence collection.  After the socket has been installed, the chamber is baked for ten days at 240~$^{\circ}$C without a trap.  We then install the trap and re-bake.  Duration and temperature of this final bakeout are limited to eight hours at 200~$^{\circ}$C followed by 24 hours at 180~$^{\circ}$C to avoid increasing trap surface roughness~\cite{AlCu}.  This bakeout is still adequate for an ion pump and a titanium sublimation pump to maintain a chamber pressure of approximately 1~$\times~10^{-9}$~Pa.

\subsection*{Trap operation}

A helical resonator (Q $\sim$ 50, driven with $\sim$0.5~W at 50-60~MHz) provides RF potentials for the trap.  Injected RF power is monitored by a capacitive divider on the resonator output, allowing for active stabilization of the applied voltage to reduce drifts in the ion secular frequencies.  The typical RF amplitude used for our measurements is approximately 200~V peak-to-peak (Vpp), but drive voltages of up to 500~Vpp or more can be tolerated without arcing.  DC potentials are supplied by National Instruments 12-bit PXI-6713 DAC cards, providing $\pm$10~V at update rates up to 500~kHz.  Simple resistor-capacitor low-pass filters with a 20~kHz 3~dB cutoff immediately outside the vacuum chamber suppress noise at the update frequency and smooth applied waveforms during ion shuttling.  We have found that these filters are inadequate to prevent heating during transport and have recently replaced them with third-order Butterworth low-pass filters (40~kHz cutoff)~\cite{BOV11e}.  Three pairs of coils compensate stray magnetic fields and create a 4~G field along $\hat{x}$ (figure~\ref{fig:trap}) to define a quantization axis.  

\subsection*{Lasers and detection optics}

Light is delivered to the vacuum chamber by single-mode optical fibers and focused across the trap surface.  Beam waists ($\sim30~\mu$m) are chosen to minimize light scatter.  Photoionization lasers (423 and 377~nm) are aligned to the loading zone along the $\hat{x}$ direction (see figure~\ref{fig:trap}) so as to minimize overlap with trap electrodes and any associated charging.  Repump lasers (866 and 854~nm) are aligned along $\hat{z}$ to uniformly illuminate the entire trapping region.  Linearly polarized beams at 397~nm propagate along $(\hat{x}-\hat{z})$ for Doppler cooling.  One of these beams is steered by a servo mirror (Optics in Motion) for computer-controlled tracking to any point along the trap.  A third $\sigma^{-}$ polarized 397~nm beam propagates along the magnetic field direction for optical pumping and state preparation.  A 729~nm laser drives the $^2$S$_{1/2}~\rightarrow~^2$D$_{5/2}$ electric quadrupole transition and is used for sideband cooling and coherent operations.




\addcontentsline{toc}{section}{References} 
\bibliographystyle{naturemag}

\bibliography{reference_database}

\begin{thebibliography}{10}
\bibitem{AlCu}
\bibinfo{note}{99\% Al / 1\% Cu has recently replaced 99\% Al / 1\% Si for
  metal layer deposition. While Al/Si is more easily etched during trap
  fabrication, ``whiskers" can grow out of the trap surface during bakeout
  which can scatter laser light or even form shorts between adjacent
  electrodes. Al/Cu metalization does not form these whiskers, though it does
  still form hillocks that can roughen the trap surface during extended bakeout
  at temperatures exceeding 200$^{\circ}$C.}
\bibitem{ionheight}
\bibinfo{note}{63~$\mu$m is 1.5 times the trap-edge waist size of a 397~nm
  laser focused to minimize light scatter, adequate to largely eliminate
  diffraction from the edge of the chip.}
\bibitem{axes}
\bibinfo{note}{Stray fields along a direction $\hat{n}$ do not in general cause
  micromotion purely along $\hat{n}$. However, in our trap the RF
  pseudopotential axes are only slightly rotated from the trap normal, so
  micromotion along the $\hat{x}$ and $\hat{y}$ directions is predominantly due
  to stray fields $E_{x}$ and $E_{y}$, respectively.}
\end{thebibliography}

\end{document}